\documentstyle[epsf,sprocl,psfig]{article}

\newcommand{\lsim}{\mathrel{\mathop{\kern 0pt \rlap
  {\raise.2ex\hbox{$<$}}}
  \lower.9ex\hbox{\kern-.190em $\sim$}}}
\newcommand{\gsim}{\mathrel{\mathop{\kern 0pt \rlap
  {\raise.2ex\hbox{$>$}}}
  \lower.9ex\hbox{\kern-.190em $\sim$}}}

\begin{document}

\title{COSMIC--RAY ANTIPROTONS FROM NEUTRALINO ANNIHILATION IN THE HALO}
\footnote{Report on  works done in collaboration with A. Bottino, F. Donato and P. Salati
and  with A. Bottino, F. Donato and S. Scopel.}

\author{Fiorenza  Donato}
\address{Dipartimento di Fisica Teorica dell'Universit\`a di Torino and 
INFN, Italy}

\maketitle\abstracts{
We report the main results of a paper where recent data of low--energy 
cosmic--ray $\bar p$ spectrum have been 
analyzed in terms of newly calculated fluxes for
secondary antiprotons and for a possible contribution of an exotic signal due
to neutralino annihilation in the galactic halo.
We also present the results of a paper in which we 
have proved that a sizeable fraction of the
supersymmetric neutralino configurations, singled out by the
DAMA/NaI data on a possible  annual modulation effect in WIMP direct search,
may provide detectable cosmic--ray antiproton signals.}

\section{Introduction}

Relic neutralinos, if present in the halo of our galaxy 
as a component of dark matter, would 
annihilate, and then produce indirect signals of various kinds. 
Among them, cosmic--ray  antiprotons are certainly 
one of the most interesting \cite{pbarnostro,pierre}
and may be detected by means of balloons or of space missions. To
discriminate this potential source of primary $\bar p$'s from the
secondary ones, we can use the different features of
their low--energy spectra ($ T_{\bar p} \lsim $ 1 GeV, $T_{\bar p}$ being
the antiproton kinetic--energy). In this energy regime the interstellar (IS)
secondary $\bar p$ spectrum
is expected to drop off very markedly because of kinematical reasons,
while primary antiprotons would show a milder fall off.
This discrimination  power is
somewhat hindered by some  effects we try 
to clarify in the following sections.

\section{Cosmic--ray proton spectrum}

We have first to fix the primary IS cosmic--ray proton spectrum, since we need
it for the evaluation of the secondary $\bar p$'s.
The IS cosmic--ray proton spectrum is derived by assuming for it appropriate
parametrizations and by fitting their corresponding solar--modulated expressions
to the TOA (top of the atmosphere) experimental fluxes.
In the present paper we use the two most recent high--statistics measurements
of the TOA proton spectrum reported  by the IMAX Collaboration 
\cite{imaxp} and by the CAPRICE Collaboration \cite{caprice}. 
We fitted these spectra using two different parametrizations:
one depending on the total proton energy, $E_p = T_p + m_p$, and the other 
on the momentum, $p$ (equivalent to rigidity for protons). 
The detailed results of our best fits to the proton data can be found in 
Table I of Ref.\cite{pierre}, in terms of the normalization
coefficient, the spectral index 
and the solar--modulation parameter $\Delta$. We find that 
even using both the parametric forms for the IS proton spectrum, the data
of the two experiments do not lead to a set of
central values for the parameters mutually compatible within their
uncertainties.  
In Fig.1 we report the median proton flux, with its uncertainty band, as
obtained from the fits to the data of the two experiments. 
\begin{figure}[h]
\vskip -20mm
\epsfxsize 110mm
\epsfysize 80mm
\centerline{\epsffile{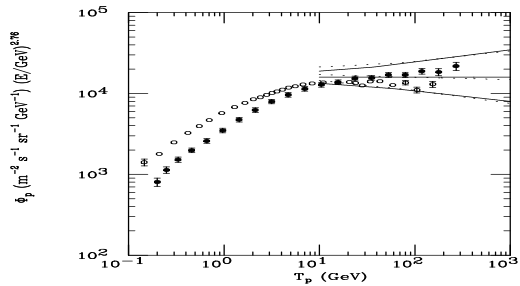}}
\vskip -20mm
\caption{TOA spectra of IMAX (full circles) and of CAPRICE (open circles). The
solid (dotted) lines denote
the median, minimal and maximal IS proton fluxes obtained with parametrization 
in energy (rigidity).}
\label{fig:fig1_idm98} 
\end{figure}
\section{Secondaries $\bar p$'s fluxes}

Cosmic ray protons interact with  hydrogen atoms at rest, lying in
the gaseus HI and HII clouds of the galactic ridge, and may
produce $\bar p$'s. This conventional spallation process is actually a
background to an hypothetical supersymmetric antiproton signal. 
The propagation of cosmic rays inside the Galaxy has been considered
in the framework of a two--zone diffusion model \cite{webber,pierre}. 
We have included energy losses in the diffusion equation, 
which tend to shift the antiproton spectrum towards lower
energies with the effect of replenishing the low--energy tail. 
The steps of the method we followed to calculate secondary  $\bar p$'s
production and diffusion are fully described in Ref.\cite{pierre}

\section{$\bar p$'s from neutralino annihilation}

The differential rate per unit volume and unit time for the production
of $ \bar p$'s from $\chi$--$\chi$ annihilation 
as a function of the kinetic energy is defined as
\begin{equation}
q_{\bar p}^{\rm susy}(T_{\bar p}) \equiv \frac{d S(T_{\bar p})}{d T_{\bar p}} =
\, <\sigma_{\rm ann} v> \, g(T_{\bar p})
\left( \frac{ \rho_\chi (r,z)}{m_\chi} \right)^2.
\label{eq:source}
\end{equation}
Here $<\sigma_{\rm ann} v>$ denotes the average over the galactic velocity
distribution function of neutralino pair annihilation cross section
$\sigma_{\rm ann}$ multiplied by the relative velocity $v$ of the annihilating
particles, $m_\chi$ is the neutralino mass and $g(T_{\bar p})$ denotes
the $\bar p$ differential spectrum. Note the dependence on the square of the
mass distribution function of neutralinos in the galactic halo, $\rho_\chi (r,z)
$. For all the details of the computation of Eq.(\ref{eq:source}) and 
the main features of the Minimal Supersymmetric Standard Model (MSSM) -- 
framework in which we calculated all the 
neutralino physical properties discussed in this talk -- 
we refer to Sect. IIIB of Ref. \cite{pierre}.\\

\section{Comparison with BESS95 data}
A recent analysis of the data collected by the BESS spectrometer during its 1995
flight \cite{bess95}
(BESS95) has provided a significant improvement in statistics 
in the kinetic--energy range
$180 ~ {\rm MeV} \leq T_{\bar{p}} \leq 1.4 ~{\rm GeV}$.
\begin{figure}[h]
\centerline{\psfig{figure=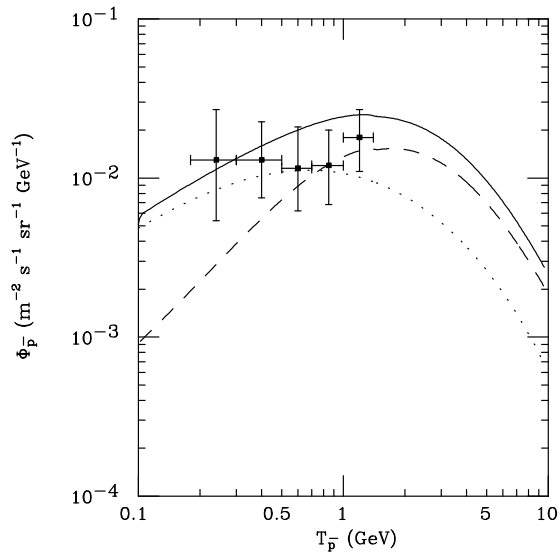,width=4.65in,bbllx=36bp,bblly=230bp,bburx=576bp,bbury=576bp,clip=}}
\caption{TOA antiproton fluxes versus the antiproton kinetic energy.The BESS95
data are shown by crosses.
The dashed line denotes the median secondary flux, the dotted one
denotes the primary flux due to neutralino annihilation in the halo for a
neutralino configuration with
$m_{\chi} = 62$ GeV, $P = 0.98$ and $\Omega_{\chi} h^2 = 0.11$.
Solid line denotes the calculated total flux.}
\label{fig:fig2_idm98} 
\end{figure}
 
From a first look at Fig.2 it is apparent that the experimental data are
rather consistent with the flux due to secondary $\bar p$'s. 
However, it is interesting to explore which would be the chances for a signal,
due to relic neutralino annihilations,
of showing  up in the low--energy window ($T_{\bar p} \lsim$ 1 GeV).
This point is very
challenging, especially in view of the interplay which might occur among
low--energy
measurements of cosmic--ray $\bar p$'s and other searches, of quite a
different nature, for relic neutralinos in our Galaxy.
Since the experimental flux seems to suggest a flatter behaviour,
as compared to the one expected for secondaries, we try to explore 
how much room for neutralino $\bar p$'s would there be in the BESS95 data. 
As a quantitative criterion to select the relevant supersymmetric
configurations, we choose to pick up only the configurations
which meet the following requirements: i) they generate a total theoretical
flux $\Phi^{\rm th}$ which is at least at the level of the experimental value
(within 1-$\sigma$) in the first energy bin; ii) their
$(\chi^2)_{\rm red}$, in the best fit of the BESS95 data, is bounded by
$(\chi^2)_{\rm red} \leq$ 2.2 (corresponding to 95\% C. L. for 5 d.o.f.).
On the other hand, supersymmetric configurations with a
$(\chi^2)_{\rm red} >$ 4 have to be considered strongly disfavoured by BESS95
data (actually, they are excluded at 99.9 \% C.L. See Ref.\cite{pierre} for
a detailed analyses of their properties).
The selected configurations are shown in
Fig.3, where $m_{\chi}$ is plotted in terms of the fractional amount
of gaugino fields, $P = a_1^2 + a_2^2$, in the neutralino mass eigenstate.
It can be seen that higgsino--like and mixed configurations are much stronger 
constrained in the neutralino mass range than the gaugino--like ones, 
because of the requirement on a rather high value of flux. 

\begin{figure}[h]
\vskip -30mm
\centerline{\psfig{figure=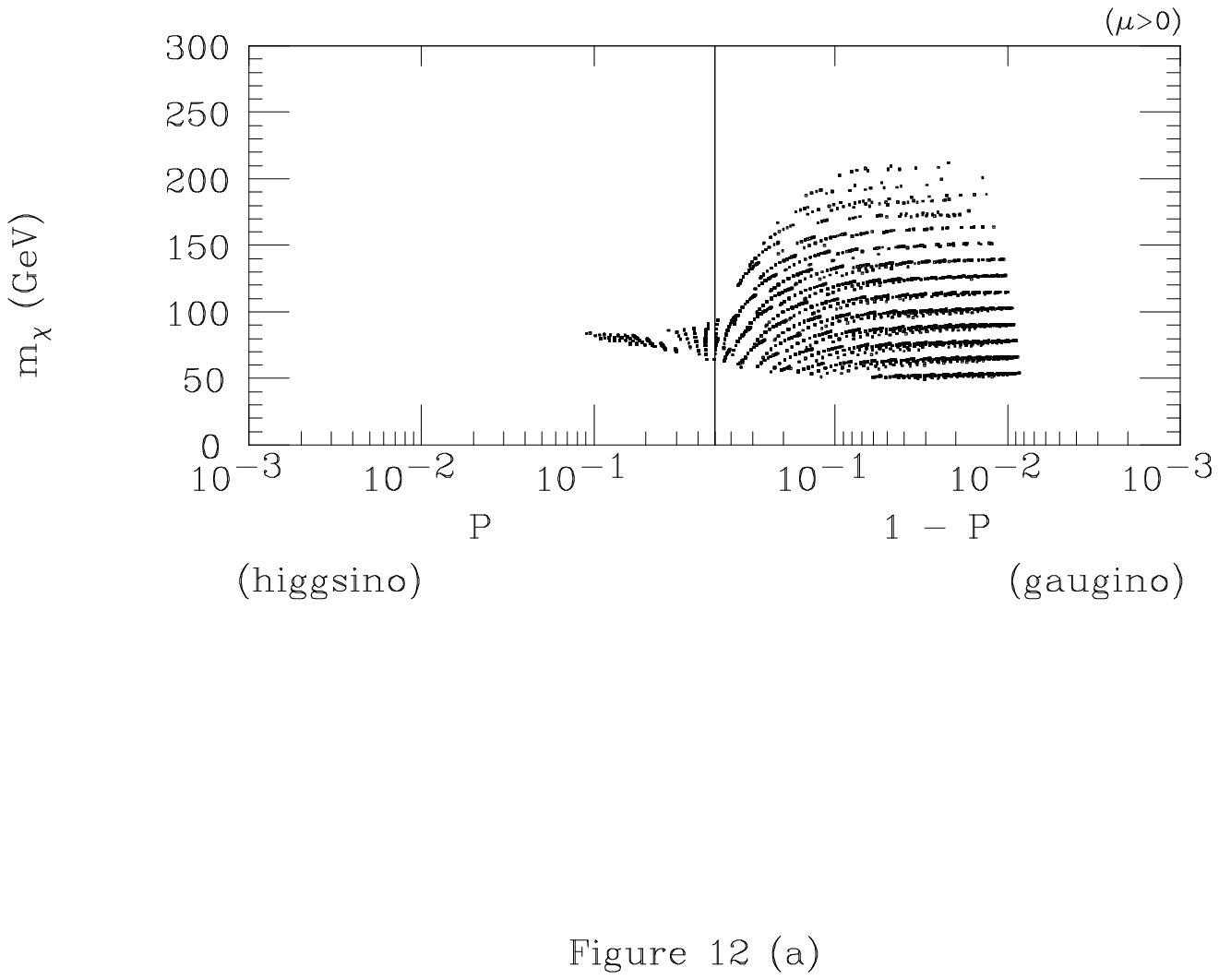,width=4.65in,bbllx=36bp,bblly=220bp,bburx=576bp,bbury=576bp,clip=}}
\vskip -20mm
\label{fig:fig3_idm98}
\caption{Scatter plots for selected (see text) configurations in the
P--$m_{\chi}$ plane. }
\end{figure}

\section{Comparison with the DAMA/NaI data on annual modulation effect}
In Refs. \cite{noi0,noi1}
we showed that the indication of  a possible
annual modulation  effect in WIMP direct search \cite{dama1,dama2}
are interpretable
in terms of a relic neutralino which may make up the major
part of dark matter in the Universe.
 
We recall that the DAMA/NaI data reported in Ref. \cite{dama2}  single out
a very delimited 2--$\sigma$ C.L. region in the plane
$\xi \sigma_{\rm scalar}^{\rm (nucleon)}$ -- $m_\chi$, where
$\sigma_{\rm scalar}^{\rm (nucleon)}$ is the WIMP--nucleon  scalar elastic
cross section  and $\xi = \rho_\chi / \rho_l$
is the fractional amount of local
WIMP density $\rho_\chi$ with respect to the total local
dark matter density $\rho_l$. 
In the analysis carried out in Ref.\cite{noi1}, 
we considered all the supersymmetric  configurations (set $S$)
which turned out to be contained in the 2--$\sigma$ C.L. region of Ref.
\cite{dama2}, by  accounting for the uncertainty in the value of $\rho_l$.

Fig.4 displays the scatter plots for TOA
antiproton fluxes calculated at $T_{\bar p} = 0.24$ GeV,
to conform to the energy range
of  the first bin of the BESS95 data (0.175 GeV $\leq T_{\bar p} \leq$ 0.3
GeV).
\begin{figure}[h]
\centerline{\psfig{figure=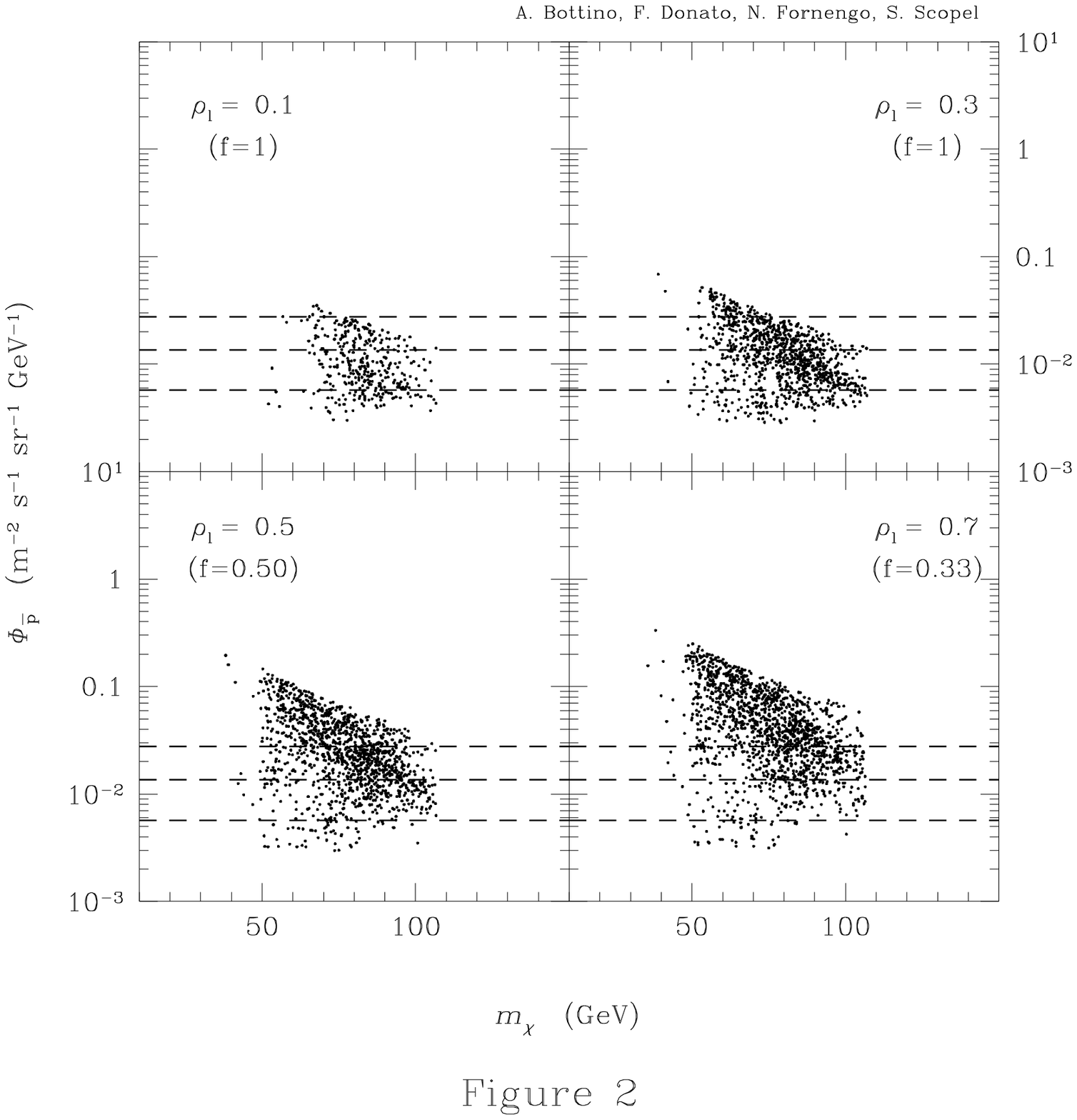,width=2.50in,bbllx=40bp,bblly=180bp,bburx=536bp,bbury=656bp,clip=}}
\label{fig:fig4_idm98} 
\caption{Scatter plots for the TOA antiproton fluxes calculated at
$T_{\bar p} = 0.24$ GeV versus the neutralino mass, for the corresponding value
of the local density $\rho_l$ and flattening factor $q$. The
dashed lines denote the central value and the 1--$\sigma$ band of the
BESS95 data in the first energy bin.}
\vskip -2mm
\end{figure}
We find that, while most of the susy configurations of the appropriate
subset of $S$ stay inside the experimental band for
$\rho_l$ = 0.1, 0.3 GeV cm$^{-3}$, at higher values of $\rho_l$ a large
number of configurations provide $\bar p$ fluxes in excess of the
experimental results. This occurrence is easily understood on the basis
of the different dependence on $\rho_l$ of the
direct detection rate and of the $\bar p$ flux, linear in the
first case and quadratic in the second one.
These results show the
remarkable property that a number  of the supersymmetric
configurations singled out by the annual modulation data  may
indeed produce
measurable effects in the low--energy part of the $\bar p$ spectrum.

We stress that the joint use of the annual modulation data in direct detection 
and of the measurements of cosmic--ray antiprotons is extremely useful 
in pinning down a number of important properties of relic neutralinos and show 
the character of complementarity of these two classes of experimental
searches for particle dark matter. This shows the great interest for
the analyses now under way of new antiproton data, those collected
by a recent balloon flight carried out by the BESS Collaboration
\cite{orito} and those measured by the AMS experiment \cite{ams}
during the June 1998 Shuttle flight.

\end{document}